\documentclass[fleqn]{2017SCGE}
\usepackage{multirow}
\usepackage{amsmath}

\begin{document}

\ensubject{subject}

\ArticleType{Article}
\Year{2020}
\Month{Jan}
\Vol{63}
\No{1}
\DOI{https://doi.org/10.1007/s11433-019-9423-3}
\ArtNo{219511}
\ReceiveDate{April 25, 2019}
\AcceptDate{April 30, 2019}
\OnlineDate{June 5, 2019}

\title{An ACE/CRIS-observation-based Galactic Cosmic Rays heavy nuclei spectra model \uppercase\expandafter{\romannumeral 2}}{An ACE/CRIS-observation-based Galactic Cosmic Rays heavy nuclei spectra model \uppercase\expandafter{\romannumeral 2}}

\author[1,2]{Shuai Fu}{}%
\author[2]{Lingling Zhao}{lz0009@uah.edu}
\author[2,3]{Gary P. Zank}{}
\author[1]{Miao Wang}{}%
\author[1]{Yong Jiang}{}

\AuthorMark{S. Fu}

\AuthorCitation{}

\address[1]{Institute of Space Weather, Nanjing University of Information Science and Technology, Nanjing, 210044, People's Republic of China;}
\address[2]{Center for Space Plasma and Aeronomic Research (CSPAR), University of Alabama in Huntsville, Huntsville, AL 35805, USA}
\address[3]{Department of Space Science, University of Alabama in Huntsville, Huntsville, AL 35899, USA}


\abstract{An observation-based galactic cosmic ray (GCR) spectral model for heavy nuclei is developed.
Zhao and Qin [J. Geophys. Res. Space Physics, 118, 1837 - 1848, 2013] proposed an empirical elemental GCR spectra model for nuclear charge $5 \leq z \leq 28$ over the energy range $\sim$30 to 500 MeV/nuc, which proved successful in predicting yearly averaged GCR heavy nuclei spectra.
Based on the latest highly statistically precise measurements from ACE/CRIS,
a further elemental GCR model with monthly averaged spectra is presented. The model can reproduce the past and predict the future
GCR intensity monthly by correlating model parameters with the
continuous sunspot number (SSN) record. The effects of solar activity on GCR modulation are considered separately for odd and even solar cycles. Compared with other comprehensive GCR models, our modeling results are satisfyingly consistent with the GCR spectral measurements from
ACE/SIS and IMP-8, and have comparable prediction accuracy as the Badhwar \& O'Neill 2014 model.
A detailed error analysis is also provided.
Finally, the GCR carbon and iron nuclei fluxes for the subsequent two solar cycles (SC 25 and 26) are predicted and they show a potential trend in reduced flux amplitude, which is suspected to be relevant to possible weak solar cycles.
}



\maketitle


\begin{multicols}{2}
\section{Introduction}\label{section1}
High-energy galactic cosmic rays (GCRs) \textcolor{black}{are} one of the major contributors to the particle radiation environment \cite{ref2}. \textcolor{black}{During the propagation of GCRs through the heliosphere}, they are modulated by the solar wind\textcolor{black}{,} and the embedded turbulent interplanetary magnetic filed (IMF)\textcolor{black}{,} and experience four major transport processes, convection, diffusion, adiabatic deceleration, and drift due to the IMF gradient and curvature \cite{ref3, ref4, ref5, ref6, ref7}. \textcolor{black}{In the process of modulation}, the GCR intensity exhibits multi-periodic features. Ground-based neutron monitors record the GCR intensity on Earth\textcolor{black}{, which peaks} every
   \Authorfootnote

\noindent 11 years and correlates inversely with \textcolor{black}{the} solar activity \cite{ref8, ref9, ref10, ref11}. During two consecutive solar minimum phases, there are alternating sharp and flat peaks of cosmic ray intensity, \textcolor{black}{which indicated} a 22-year (yr) solar cycle variation (the so-called Hale cycle \cite{ref12}). This 22-yr cycle is considered to be related to the polarity reversal of the IMF \cite{ref13}. In addition, a 27-day periodic change of the GCR intensity is also observed, which is thought to be caused by the 27-day variations of the solar wind velocity \cite {ref14, ref15, ref16}.

GCRs consist of $\sim85\%$ protons, $\sim12\%$ alpha particles, and $\sim3\%$ heavy nuclei ($z\geq3$) and electrons \cite{ref17}. Although the abundance of heavy nuclei is relatively low, they can play key roles in geophysical phenomena due to their large nucleon numbers and high energies \cite{ref18}. Heavy nuclei with energies above 30 MeV/nuc account for approximately half of the radiation dose in the GCR environment, and the relatively abundant heavy ions (e.g., C, N, O, Fe) are the dominant cause \cite{ref19, ref20}. It is\textcolor{black}{, therefore,} necessary to account for \textcolor{black}{the heavy} GCR ions in evaluating the space radiation environment. Numerous models have been developed to study GCR protons \cite{ref21, ref22, ref7, ref23}. However, a comprehensive understanding of the \textcolor{black}{heavy} GCR ions spectra and \textcolor{black}{the} abundances remains challenging partially due to the relatively low intensities and the lack of measurements \cite{ref24, ref25, ref26, ref27, ref28, ref10}.

The derivation of commonly used GCR spectral models falls into two categories\textcolor{black}{;} analyzing and fitting \textcolor{black}{the} experimental data (empirical model)\textcolor{black}{,} and numerically solving the \textcolor{black}{Parker} transport equation (physical model) \cite{ref29}. Physical models consider GCRs transport processes throughout the heliosphere and the induced effects. Many GCR models have been \textcolor{black}{developed} and compared with observations from spacecrafts and ground-based neutron monitors during the past several decades, and the model uncertainty is often claimed to be within 15\% \cite{ref30, ref31}. However, results from equation-based physical models cannot be directly applied to estimate radiation exposure in the absence of adequate experimental-based corrections. From this perspective, empirical models are frequently adopted to evaluate the space radiation hazard \cite{ref32, ref33, ref18}. The Cosmic Ray Effects on Micro-Electronics (CR$\grave{E}$ME) model \cite{ref34, ref35, ref36} and the NASA Badhwar-O'Neill (BON) model \cite{ref37, ref38, ref39, ref40, ref41} are two popular models\textcolor{black}{, that are} used to provide accurate GCR spectra near the Earth. The CR$\grave{E}$ME model was developed initially to calculate the effect of interplanetary space radiation on electronic systems, as first proposed by the Naval Research Laboratory \cite{ref34}. The CR$\grave{E}$ME model relates solar cycle variation in the GCR intensity with the historical sunspot number \cite{ref42}. By using sunspot number as a representation of solar modulation, the CR$\grave{E}$ME2009 model offers more advantages than \textcolor{black}{the }other GCR models in predicting the modulation level up to one year in advance \cite{ref36}. Unlike the CR$\grave{E}$ME model that describes the GCR particle fluxes semi-empirically, the BON model numerically solves the propagation equation in the heliosphere. In \textcolor{black}{the} BON model, solar modulation is derived from the modulation potential $\Phi(t)$, which is a time-dependent parameter and is a function of particle rigidity \cite{ref38}.  It is reported that BON2010 and BON2011 show large offsets when modeling GCR fluxes near \textcolor{black}{the} solar minimum 2010 and maximum 2000 \cite{ref43}. BON2011 systematically tends to overestimate the fluxes of heavier nuclei ($z\geq3$) within the energy 0.5-4 GeV/nucleon. BON2014 is the latest model, which focuses more attention on GCRs at higher energies\textcolor{black}{, that are} not detected by ACE/CRIS, and has been proved to be a substantial improvement over previous versions \cite{ref41}. A comparison between three widely used GCR models (CR$\grave{E}$ME96, CR$\grave{E}$ME2009 and Badhwar-O'Neill 2010) was conducted by Mrigakshi et al. \cite{ref44} to investigate their accuracy in estimating radiation exposure. They concluded that these models cannot perfectly describe the changes in GCR intensity resulting from solar modulation, and large discrepancies between the model results and the measurements can be seen within a given period. In addition, many commonly used models attach great importance to particles with energies higher than GeV/nuc, which do not contribute significantly to most radiation exposure\textcolor{black}{,} and atmospheric processes because of the lower fluxes \cite{ref18}. The effect of GCRs with lower energies (several tens \textcolor{black}{of} MeV to 1 GeV) is usually underestimated.

Zhao and Qin 2013 model (henceforth ZQ13) presented an observation-based elemental GCR spectral model for heavy nuclei \cite{ref1} . With four free parameters, their model is in good agreement with the observed GCR spectra in the energy range 30--500 MeV/nuc. By relating the model parameters to the annual averaged sunspot numbers (SSNs), the ZQ13 model can reconstruct yearly GCR energy spectra for earlier periods and predict the spectra for the subsequent solar cycle. However, the ZQ13 model is based on a yearly averaged heavy nuclei experiment and its temporal precision for prediction is limited. In this paper, we present a refined GCR heavy nuclei spectra model using monthly averaged intensity measurements from ACE/CRIS. The time lag between SSN and the GCR intensity level near Earth is considered separately for odd and even solar cycles. The paper is organized as follows. In Section 2, we present the dataset used to build the model. In Section 3, the development of the model is presented briefly. In Section 4, modeling results are compared with measurements and other well-known GCR models.

\begin{table*}[t]
  \centering
  \footnotesize
  \caption{The measured energy bands from ACE/CRIS instrument}
    \begin{tabular}{ccccccccccccccc}
 \toprule
    \multicolumn{1}{l}{Element($z$)} & \multicolumn{1}{c}{$\emph{E}_{1}$} & \multicolumn{1}{c}{$\Delta \emph{E}_{1}$} & \multicolumn{1}{c}{$\emph{E}_{2}$} & \multicolumn{1}{c}{$\Delta \emph{E}_{2}$} & \multicolumn{1}{c}{$\emph{E}_{3}$} & \multicolumn{1}{c}{$\Delta \emph{E}_{3}$} & \multicolumn{1}{c}{$\emph{E}_{4}$} & \multicolumn{1}{c}{$\Delta \emph{E}_{4}$} & \multicolumn{1}{c}{$\emph{E}_{5}$} & \multicolumn{1}{c}{$\Delta \emph{E}_{5}$} & \multicolumn{1}{c}{$\emph{E}_{6}$} & \multicolumn{1}{c}{$\Delta \emph{E}_{6}$} & \multicolumn{1}{c}{$\emph{E}_{7}$} & \multicolumn{1}{c}{$\Delta \emph{E}_{7}$} \\
    \midrule
\multicolumn{1}{l}{B(5)} & \multicolumn{1}{c}{59.6} & \multicolumn{1}{c}{14.4} & \multicolumn{1}{c}{79.7} & \multicolumn{1}{c}{23.1} & \multicolumn{1}{c}{102.0} & \multicolumn{1}{c}{19.1} & \multicolumn{1}{c}{121.1} & \multicolumn{1}{c}{16.8} & \multicolumn{1}{c}{138.2} & \multicolumn{1}{c}{15.3} & \multicolumn{1}{c}{154.0} & \multicolumn{1}{c}{14.1} & \multicolumn{1}{c}{168.6} & \multicolumn{1}{c}{13.5} \\
    \multicolumn{1}{l}{C(6)} & \multicolumn{1}{c}{68.3} & \multicolumn{1}{c}{16.6} & \multicolumn{1}{c}{91.5} & \multicolumn{1}{c}{26.6} & \multicolumn{1}{c}{117.3} & \multicolumn{1}{c}{22.1} & \multicolumn{1}{c}{139.3} & \multicolumn{1}{c}{19.5} & \multicolumn{1}{c}{159.1} & \multicolumn{1}{c}{17.8} & \multicolumn{1}{c}{177.4} & \multicolumn{1}{c}{16.4} & \multicolumn{1}{c}{194.5} & \multicolumn{1}{c}{15.7} \\
    \multicolumn{1}{l}{N(7)} & \multicolumn{1}{c}{73.3} & \multicolumn{1}{c}{17.8} & \multicolumn{1}{c}{98.1} & \multicolumn{1}{c}{28.5} & \multicolumn{1}{c}{125.9} & \multicolumn{1}{c}{23.8} & \multicolumn{1}{c}{149.6} & \multicolumn{1}{c}{21.0} & \multicolumn{1}{c}{171.0} & \multicolumn{1}{c}{19.2} & \multicolumn{1}{c}{190.7} & \multicolumn{1}{c}{17.7} & \multicolumn{1}{c}{209.2} & \multicolumn{1}{c}{16.9} \\
    \multicolumn{1}{l}{O(8)} & \multicolumn{1}{c}{80.4} & \multicolumn{1}{c}{19.6} & \multicolumn{1}{c}{107.8} & \multicolumn{1}{c}{31.5} & \multicolumn{1}{c}{138.4} & \multicolumn{1}{c}{26.3} & \multicolumn{1}{c}{164.7} & \multicolumn{1}{c}{23.3} & \multicolumn{1}{c}{188.4} & \multicolumn{1}{c}{21.3} & \multicolumn{1}{c}{210.3} & \multicolumn{1}{c}{19.7} & \multicolumn{1}{c}{230.8} & \multicolumn{1}{c}{18.8} \\
    \multicolumn{1}{l}{F(9)} & \multicolumn{1}{c}{83.5} & \multicolumn{1}{c}{20.4} & \multicolumn{1}{c}{112.0} & \multicolumn{1}{c}{32.8} & \multicolumn{1}{c}{143.8} & \multicolumn{1}{c}{27.4} & \multicolumn{1}{c}{171.1} & \multicolumn{1}{c}{24.3} & \multicolumn{1}{c}{195.9} & \multicolumn{1}{c}{22.2} & \multicolumn{1}{c}{218.7} & \multicolumn{1}{c}{20.6} & \multicolumn{1}{c}{240.0} & \multicolumn{1}{c}{19.6} \\
    \multicolumn{1}{l}{Ne(10)} & \multicolumn{1}{c}{89.5} & \multicolumn{1}{c}{21.8} & \multicolumn{1}{c}{120.1} & \multicolumn{1}{c}{35.2} & \multicolumn{1}{c}{154.4} & \multicolumn{1}{c}{29.5} & \multicolumn{1}{c}{183.9} & \multicolumn{1}{c}{26.1} & \multicolumn{1}{c}{210.6} & \multicolumn{1}{c}{24.0} & \multicolumn{1}{c}{235.3} & \multicolumn{1}{c}{22.2} & \multicolumn{1}{c}{258.4} & \multicolumn{1}{c}{21.2} \\
    \multicolumn{1}{l}{Na(11)} & \multicolumn{1}{c}{94.0} & \multicolumn{1}{c}{23.1} & \multicolumn{1}{c}{126.2} & \multicolumn{1}{c}{37.1} & \multicolumn{1}{c}{162.4} & \multicolumn{1}{c}{31.2} & \multicolumn{1}{c}{193.5} & \multicolumn{1}{c}{27.6} & \multicolumn{1}{c}{221.7} & \multicolumn{1}{c}{25.4} & \multicolumn{1}{c}{247.8} & \multicolumn{1}{c}{23.5} & \multicolumn{1}{c}{272.3} & \multicolumn{1}{c}{22.6} \\
    \multicolumn{1}{l}{Mg(12)} & \multicolumn{1}{c}{100.2} & \multicolumn{1}{c}{24.6} & \multicolumn{1}{c}{134.7} & \multicolumn{1}{c}{39.8} & \multicolumn{1}{c}{173.4} & \multicolumn{1}{c}{33.3} & \multicolumn{1}{c}{206.8} & \multicolumn{1}{c}{29.6} & \multicolumn{1}{c}{237.1} & \multicolumn{1}{c}{27.3} & \multicolumn{1}{c}{265.2} & \multicolumn{1}{c}{25.4} & \multicolumn{1}{c}{291.5} & \multicolumn{1}{c}{24.2} \\
    \multicolumn{1}{l}{Al(13)} & \multicolumn{1}{c}{103.8} & \multicolumn{1}{c}{25.6} & \multicolumn{1}{c}{139.6} & \multicolumn{1}{c}{41.3} & \multicolumn{1}{c}{179.8} & \multicolumn{1}{c}{34.7} & \multicolumn{1}{c}{214.5} & \multicolumn{1}{c}{30.9} & \multicolumn{1}{c}{246.1} & \multicolumn{1}{c}{28.4} & \multicolumn{1}{c}{275.3} & \multicolumn{1}{c}{26.4} & \multicolumn{1}{c}{302.8} & \multicolumn{1}{c}{25.2} \\
    \multicolumn{1}{l}{Si(14)} & \multicolumn{1}{c}{110.1} & \multicolumn{1}{c}{27.1} & \multicolumn{1}{c}{148.2} & \multicolumn{1}{c}{44.0} & \multicolumn{1}{c}{191.1} & \multicolumn{1}{c}{37.0} & \multicolumn{1}{c}{228.1} & \multicolumn{1}{c}{33.0} & \multicolumn{1}{c}{261.8} & \multicolumn{1}{c}{30.4} & \multicolumn{1}{c}{293.1} & \multicolumn{1}{c}{28.3} & \multicolumn{1}{c}{322.6} & \multicolumn{1}{c}{27.1} \\
    \multicolumn{1}{l}{P(15)} & \multicolumn{1}{c}{112.7} & \multicolumn{1}{c}{27.8} & \multicolumn{1}{c}{151.8} & \multicolumn{1}{c}{45.1} & \multicolumn{1}{c}{195.9} & \multicolumn{1}{c}{38.0} & \multicolumn{1}{c}{233.9} & \multicolumn{1}{c}{33.9} & \multicolumn{1}{c}{268.6} & \multicolumn{1}{c}{31.2} & \multicolumn{1}{c}{300.8} & \multicolumn{1}{c}{29.1} & \multicolumn{1}{c}{331.1} & \multicolumn{1}{c}{27.9} \\
    \multicolumn{1}{l}{S(16)} & \multicolumn{1}{c}{118.2} & \multicolumn{1}{c}{29.2} & \multicolumn{1}{c}{159.4} & \multicolumn{1}{c}{47.4} & \multicolumn{1}{c}{205.8} & \multicolumn{1}{c}{40.1} & \multicolumn{1}{c}{245.9} & \multicolumn{1}{c}{35.8} & \multicolumn{1}{c}{282.5} & \multicolumn{1}{c}{32.9} & \multicolumn{1}{c}{316.6} & \multicolumn{1}{c}{30.7} & \multicolumn{1}{c}{348.7} & \multicolumn{1}{c}{29.5} \\
    \multicolumn{1}{l}{Cl(17)} & \multicolumn{1}{c}{120.2} & \multicolumn{1}{c}{29.8} & \multicolumn{1}{c}{162.1} & \multicolumn{1}{c}{48.2} & \multicolumn{1}{c}{209.4} & \multicolumn{1}{c}{40.8} & \multicolumn{1}{c}{250.3} & \multicolumn{1}{c}{36.4} & \multicolumn{1}{c}{287.7} & \multicolumn{1}{c}{33.6} & \multicolumn{1}{c}{322.4} & \multicolumn{1}{c}{31.4} & \multicolumn{1}{c}{355.1} & \multicolumn{1}{c}{30.1} \\
    \multicolumn{1}{l}{Ar(18)} & \multicolumn{1}{c}{125.0} & \multicolumn{1}{c}{31.1} & \multicolumn{1}{c}{168.8} & \multicolumn{1}{c}{50.5} & \multicolumn{1}{c}{218.1} & \multicolumn{1}{c}{42.7} & \multicolumn{1}{c}{260.9} & \multicolumn{1}{c}{38.2} & \multicolumn{1}{c}{300.0} & \multicolumn{1}{c}{35.3} & \multicolumn{1}{c}{336.4} & \multicolumn{1}{c}{32.9} & \multicolumn{1}{c}{370.8} & \multicolumn{1}{c}{31.7} \\
    \multicolumn{1}{l}{K(19)} & \multicolumn{1}{c}{127.9} & \multicolumn{1}{c}{31.9} & \multicolumn{1}{c}{172.8} & \multicolumn{1}{c}{51.7} & \multicolumn{1}{c}{223.4} & \multicolumn{1}{c}{43.9} & \multicolumn{1}{c}{267.4} & \multicolumn{1}{c}{39.2} & \multicolumn{1}{c}{307.5} & \multicolumn{1}{c}{36.3} & \multicolumn{1}{c}{344.9} & \multicolumn{1}{c}{33.9} & \multicolumn{1}{c}{380.3} & \multicolumn{1}{c}{32.5} \\
    \multicolumn{1}{l}{Ca(20)} & \multicolumn{1}{c}{131.6} & \multicolumn{1}{c}{32.8} & \multicolumn{1}{c}{177.9} & \multicolumn{1}{c}{53.5} & \multicolumn{1}{c}{230.1} & \multicolumn{1}{c}{45.3} & \multicolumn{1}{c}{275.6} & \multicolumn{1}{c}{40.5} & \multicolumn{1}{c}{317.1} & \multicolumn{1}{c}{37.5} & \multicolumn{1}{c}{355.9} & \multicolumn{1}{c}{35.1} & \multicolumn{1}{c}{392.4} & \multicolumn{1}{c}{33.8} \\
    \multicolumn{1}{l}{Sc(21)} & \multicolumn{1}{c}{133.5} & \multicolumn{1}{c}{33.4} & \multicolumn{1}{c}{180.5} & \multicolumn{1}{c}{54.4} & \multicolumn{1}{c}{233.7} & \multicolumn{1}{c}{46.0} & \multicolumn{1}{c}{279.9} & \multicolumn{1}{c}{41.2} & \multicolumn{1}{c}{322.2} & \multicolumn{1}{c}{38.1} & \multicolumn{1}{c}{361.6} & \multicolumn{1}{c}{35.7} & \multicolumn{1}{c}{398.8} & \multicolumn{1}{c}{34.4} \\
    \multicolumn{1}{l}{Ti(22)} & \multicolumn{1}{c}{137.1} & \multicolumn{1}{c}{34.3} & \multicolumn{1}{c}{185.5} & \multicolumn{1}{c}{55.9} & \multicolumn{1}{c}{240.3} & \multicolumn{1}{c}{47.5} & \multicolumn{1}{c}{287.9} & \multicolumn{1}{c}{42.5} & \multicolumn{1}{c}{331.6} & \multicolumn{1}{c}{39.4} & \multicolumn{1}{c}{372.3} & \multicolumn{1}{c}{36.9} & \multicolumn{1}{c}{410.8} & \multicolumn{1}{c}{35.5} \\
    \multicolumn{1}{l}{V(23)} & \multicolumn{1}{c}{139.9} & \multicolumn{1}{c}{35.1} & \multicolumn{1}{c}{189.5} & \multicolumn{1}{c}{57.2} & \multicolumn{1}{c}{245.5} & \multicolumn{1}{c}{48.6} & \multicolumn{1}{c}{294.3} & \multicolumn{1}{c}{43.6} & \multicolumn{1}{c}{339.1} & \multicolumn{1}{c}{40.4} & \multicolumn{1}{c}{380.8} & \multicolumn{1}{c}{37.9} & \multicolumn{1}{c}{420.3} & \multicolumn{1}{c}{36.4} \\
    \multicolumn{1}{l}{Cr(24)} & \multicolumn{1}{c}{144.0} & \multicolumn{1}{c}{36.1} & \multicolumn{1}{c}{195.1} & \multicolumn{1}{c}{59.0} & \multicolumn{1}{c}{253.0} & \multicolumn{1}{c}{50.2} & \multicolumn{1}{c}{303.5} & \multicolumn{1}{c}{45.0} & \multicolumn{1}{c}{349.8} & \multicolumn{1}{c}{41.8} & \multicolumn{1}{c}{393.0} & \multicolumn{1}{c}{39.1} & \multicolumn{1}{c}{434.0} & \multicolumn{1}{c}{37.8} \\
    \multicolumn{1}{l}{Mn(25)} & \multicolumn{1}{c}{146.8} & \multicolumn{1}{c}{37.0} & \multicolumn{1}{c}{199.1} & \multicolumn{1}{c}{60.3} & \multicolumn{1}{c}{258.3} & \multicolumn{1}{c}{51.3} & \multicolumn{1}{c}{309.9} & \multicolumn{1}{c}{46.2} & \multicolumn{1}{c}{357.3} & \multicolumn{1}{c}{42.8} & \multicolumn{1}{c}{401.6} & \multicolumn{1}{c}{40.2} & \multicolumn{1}{c}{443.5} & \multicolumn{1}{c}{38.7} \\
    \multicolumn{1}{l}{Fe(26)} & \multicolumn{1}{c}{150.4} & \multicolumn{1}{c}{37.9} & \multicolumn{1}{c}{204.1} & \multicolumn{1}{c}{62.1} & \multicolumn{1}{c}{265.0} & \multicolumn{1}{c}{52.8} & \multicolumn{1}{c}{318.1} & \multicolumn{1}{c}{47.5} & \multicolumn{1}{c}{366.9} & \multicolumn{1}{c}{44.1} & \multicolumn{1}{c}{412.6} & \multicolumn{1}{c}{41.4} & \multicolumn{1}{c}{455.9} & \multicolumn{1}{c}{39.9} \\
    \multicolumn{1}{l}{Co(27)} & \multicolumn{1}{c}{153.6} & \multicolumn{1}{c}{38.9} & \multicolumn{1}{c}{208.5} & \multicolumn{1}{c}{63.4} & \multicolumn{1}{c}{270.9} & \multicolumn{1}{c}{54.1} & \multicolumn{1}{c}{325.3} & \multicolumn{1}{c}{48.7} & \multicolumn{1}{c}{375.4} & \multicolumn{1}{c}{45.2} & \multicolumn{1}{c}{422.3} & \multicolumn{1}{c}{42.5} & \multicolumn{1}{c}{466.7} & \multicolumn{1}{c}{41.1} \\
    \multicolumn{1}{l}{Ni(28)} & \multicolumn{1}{c}{158.9} & \multicolumn{1}{c}{40.2} & \multicolumn{1}{c}{215.9} & \multicolumn{1}{c}{66.0} & \multicolumn{1}{c}{280.7} & \multicolumn{1}{c}{56.4} & \multicolumn{1}{c}{337.3} & \multicolumn{1}{c}{50.8} & \multicolumn{1}{c}{389.5} & \multicolumn{1}{c}{47.1} & \multicolumn{1}{c}{438.4} & \multicolumn{1}{c}{44.4} & \multicolumn{1}{c}{484.7} & \multicolumn{1}{c}{42.9} \\
   \bottomrule
\end{tabular}
\begin{threeparttable}[b]
\begin{tablenotes}
  \item[*] $\emph{E}_{i}$ is the recommended midpoint of each energy band, and $\Delta \emph{E}_{i}$ is the corresponding energy band. All energies have a unit of MeV/nuc.
\end{tablenotes}
\end{threeparttable}
  \label{table1}%
\end{table*}

\section{Dataset}\label{sec:2}
The NASA Advanced Composition Explorer (ACE), launched on August 25, 1997, has collected continuously a large amount of solar wind, interplanetary magnetic field, solar energetic particles and cosmic ray data \cite{ref45}. The ACE Science Center (ASC) ensures those observations are well documented and publicly available. One of the primary instruments onboard ACE is the Cosmic Ray Isotope Spectrometer (CRIS), which measures the intensities of GCR heavy ions from boron ($z=5$) to nickel ($z=28$) with kinetic energies between $\sim$50 to $\sim$500 MeV/nuc. The energy range is specified \textcolor{black}{in} Table \ref{table1}. The Solar Isotope Spectrometer (SIS) instrument onboard ACE provides high resolution measurements of the isotopic composition of 14 energetic nuclei (He, C, N, O, Ne, Na, Mg, Al, Si, S, Ar, Ca, Fe, and Ni) covering the energy range \textcolor{black}{from} $\sim$10 to $\sim$100 MeV/nuc, solar energetic particles during large solar eruptive events, and galactic and anomalous cosmic rays as a background. A flag for each measurement is provided to distinguish the solar activity level with $flag = 0$ for \textcolor{black}{quiet solar condition} and $flag = 1$ for \textcolor{black}{active solar} conditions.

To develop our GCR model, the daily averaged differential fluxes (in units of particles/($cm^{2} \cdot sr \cdot s \cdot MeV/nuc$), species from boron to nickel) from ACE/CRIS and SIS instruments are collected \textcolor{black}{through} January 1998 to October 2018, from the New ACE Level 2 Data Server (\url{http://www.srl.caltech.edu/ACE/ASC/level2/new/intro.html}). Data between August \textcolor{black}{and} December, 1997 is not used to build the model due to \textcolor{black}{the} issues related to \textcolor{black}{the} calibration as referenced \textcolor{black}{in} the website \url{http://www.srl.caltech.edu/ACE/ASC/level2/cris_l2desc.html}. As for SIS, only observations during \textcolor{black}{quiet solar} conditions ($flag = 0$) are used. The daily data is \textcolor{black}{monthly averaged}.

Before the launch of ACE in 1997, IMP-8 was recording cosmic ray flux over a long period. To compare with these earlier data, we digitize the 27-day averaged IMP-8 C, N, O and Fe flux during the period 1975--1995 \textcolor{black}{(see Figure 1 of reference \cite{ref35})}. Additionally, we also compare our modeled results with other well-known GCR models, i.e., the CR$\grave{E}$ME2009 model, which is available online \url{https://creme.isde.vanderbilt.edu/}. It should be noted that CR$\grave{E}$ME2009 \textcolor{black}{provides} daily predictions. We pick out 10 days randomly in each month and simulate GCR fluxes online. By averaging over the 10 days, the monthly GCR predictions of CR$\grave{E}$ME2009 are obtained. Because the GCR flux provided by the CR$\grave{E}$ME model does not change significantly in such a short time interval \cite{ref44}, the processing method is meaningful and acceptable. Part of the simulation results from BON2010, 2011 and 2014 models are found from other literature \cite{ref44, ref41}.

As a proxy of solar activity, the monthly averaged international Sunspot Number (SSN) (V2.0) (\url{http://www.sidc.be/silso/datafiles}) is used in this work to study the relationship between GCR model parameters and solar activity.

\section{Model Description}\label{sec:3}

As presented in the derivation of the ZQ13 model, the development of a GCR spectral model is divided roughly into three parts: an integral intensity model $I^{m}(z,t)$, a spectral shape model $g(E,t)$, and finally an energy spectra model $f(z,t,E)$. To avoid repetition, we present a general description about the procedures.

\subsection{Integral Intensity Model}

\textcolor{black}{We} define the integral intensity $I(z,t)$ \textcolor{black}{as,}
\begin{center}
\begin{equation}
I(z,t) = \sum_{i=1}^{7} f_{i}(z,t) \Delta E_{i}\label{eq1},
\end{equation}
\end{center}
where $f_{i}(z,t)$ represents the monthly averaged differential flux \textcolor{black}{of} element $z$ at time $t$, and $\Delta E_{i}$ is the $i$th energy interval listed in Table \ref{table1}. It would seem that the integral intensity is related to elemental abundance, but elemental abundance is derived by integrating over a fixed energy band while the ACE/CRIS energy range is species specified. So, there is no comparability between the two quantities.

The intensity ratio of each element relative to oxygen ($z=8$) is defined as the Integral Intensity Ratio (IIR) $Z(z,t)$,
\begin{center}
\begin{equation}
Z(z,t) = \frac{I(z,t)}{I(z=8,t)}\label{eq2}.
\end{equation}
\end{center}

The averaged intensity ratio $\bar{Z}(z)$ is defined as
\begin{center}
\begin{equation}
\bar{Z}(z) = \frac{1}{N} \sum_{t=1}^{N} Z(z,t)\label{eq3},
\end{equation}
\end{center}
where $N$\textcolor{black}{(=250)} is the number of months used for developing our model.

The integral intensity ratio percentage (IRP) $p(z,t)$, the elemental-averaged IRP $\bar{p}(t)$, and the ratio between them are separately defined as:
\begin{center}
\begin{equation}
p(z,t) = \frac{Z(z,t)-\bar{Z}(z)}{\sqrt{\bar{Z}(z)}} \times 100\%\label{eq4},
\end{equation}
\end{center}

\begin{center}
\begin{equation}
\bar{p}(t) = \frac{1}{24} \sum_{z=5}^{28} p(z,t)\label{eq5},
\end{equation}
\end{center}

\begin{center}
\begin{equation}
\lambda(z,t) = \frac{p(z,t)}{\bar{p}(t)}\label{eq6},
\end{equation}
\end{center}

In \cite{ref1}, they set the $\lambda$ of C(6) as zero, the $\lambda$ of Fe(26) as 3.3, and $\lambda$ as 1 for the other elements, which means $p(z,t) = \bar{p}(t)$ for most nuclei. This corresponds to the second assumption in ZQ13 model that the intensity radio percentages (IRP) for all the elements, except C(6), O(8) and Fe(26), are the same with the elemental average. In other words, $\lambda(z,t)\equiv1$ (except $z=6,8,26$). However, such an approximation of $\lambda(z)$ is not precise enough, \textcolor{black}{so that it} may introduce some discrepancies. In this paper, a refined parameter is \textcolor{black}{calculated as,}

\begin{center}
\begin{equation}
\lambda^*(z) = \frac{1}{N} \sum_{t=1}^{N} \lambda(z,t)\label{eq7}.
\end{equation}
\end{center}

From Equations \eqref{eq4}-\eqref{eq7}, the integral intensity ratio model can be derived as,
\begin{center}
\begin{equation}
Z^{m}(z,t) = \bar{Z}(z) \left( 1+ \frac{p(z,t)}{\sqrt{\bar{Z}(z)}} \right) = \bar{Z}(z) \left( 1+ \frac{\lambda^*(z) \bar{p}(t)}{\sqrt{\bar{Z}(z)}} \right)\label{eq8},
\end{equation}
\end{center}
where $Z^{m}(z,t)$ is the modeled integral intensity ratio for element $z$ at arbitrary time $t$, and $\bar{Z}(z)$ and $\bar{p}(t)$ represent the time-averaged intensity ratio and elemental-averaged intensity ratio percentage, respectively. We use the actual 24 values of $\lambda^*(z)$ to represent the 24 elements from boron to nickel (Table \ref{table2}). Although the $\lambda(z)$ values used in ZQ13 model are relatively close to $\lambda^*(z)$, it is still better to use the actual parameter values instead of approximating, in order to ensure maximum accuracy of the model.

\begin{table*}[t]
  \centering
  \footnotesize
  \caption{Parameter $\lambda^*(z)$ used in our GCR model}
    \begin{tabular}{rrrrrrrrrrrrr}
   \toprule
    \multicolumn{1}{l}{$z$} & \multicolumn{1}{c}{5} & \multicolumn{1}{c}{6} & \multicolumn{1}{c}{7} & \multicolumn{1}{c}{8} & \multicolumn{1}{c}{9} & \multicolumn{1}{c}{10} & \multicolumn{1}{c}{11} & \multicolumn{1}{c}{12} & \multicolumn{1}{c}{13} & \multicolumn{1}{c}{14} & \multicolumn{1}{c}{15} & \multicolumn{1}{c}{16} \\
    \midrule
    \multicolumn{1}{l}{$\lambda(z)$} & \multicolumn{1}{c}{0.77} & \multicolumn{1}{c}{0.12} & \multicolumn{1}{c}{0.77} & \multicolumn{1}{c}{0.00} & \multicolumn{1}{c}{0.55} & \multicolumn{1}{c}{1.01} & \multicolumn{1}{c}{0.54} & \multicolumn{1}{c}{1.44} & \multicolumn{1}{c}{1.09} & \multicolumn{1}{c}{1.59} & \multicolumn{1}{c}{0.73} & \multicolumn{1}{c}{1.34} \\
\bottomrule
 \toprule
    \multicolumn{1}{l}{$z$} & \multicolumn{1}{c}{17} & \multicolumn{1}{c}{18} & \multicolumn{1}{c}{19} & \multicolumn{1}{c}{20} & \multicolumn{1}{c}{21} & \multicolumn{1}{c}{22} & \multicolumn{1}{c}{23} & \multicolumn{1}{c}{24} & \multicolumn{1}{c}{25} & \multicolumn{1}{c}{26} & \multicolumn{1}{c}{27} & \multicolumn{1}{c}{28} \\
    \midrule
    \multicolumn{1}{l}{$\lambda(z)$} & \multicolumn{1}{c}{0.96} & \multicolumn{1}{c}{1.29} & \multicolumn{1}{c}{1.02} & \multicolumn{1}{c}{1.42} & \multicolumn{1}{c}{0.75} & \multicolumn{1}{c}{1.56} & \multicolumn{1}{c}{0.81} & \multicolumn{1}{c}{1.27} & \multicolumn{1}{c}{1.04} & \multicolumn{1}{c}{3.38} & \multicolumn{1}{c}{0.33} & \multicolumn{1}{c}{1.15} \\
      \bottomrule
          &       &       &       &       &       &       &       &       &       &       &       &  \\
    \end{tabular}%
  \label{table2}%
\end{table*}%

To model the effect of solar modulation on GCR intensity, the intensity modulation parameter, $\alpha(t)$, is introduced. For each month, using a linear relationship between \textcolor{black}{the} atomic number ($z$) and \textcolor{black}{the} logarithmic $I^{c}(z,t)$, $\alpha (t)$ is given by
\begin{center}
\begin{equation}
log10[I^{c}(z,t)]\approx \alpha +\beta z\label{eq9},
\end{equation}
\end{center}
where, $I^{c}(z,t)$ is the corrected integral intensity calculated by
\begin{center}
\begin{equation}
I^{c}(z,t) = \frac{I(z,t)}{Z^{m}(z,t)}\label{eq10}.
\end{equation}
\end{center}

On combining the integral intensity ratio model $Z^{m}(z,t)$ with the intensity modulation parameter $\alpha(t)$, the integral intensity model can be derived as
\begin{center}
\begin{equation}
 I^{m}(z,t)=Z^{m}(z,t)10^{\alpha (t)}\label{eq11}.
 \end{equation}
\end{center}

\subsection{Spectral Shape Function}
In the ZQ13 model, the fundamental assumption is that all elements have the same spectral shape at the same time. We adopt this assumption and use Equation \eqref{eq12}  to fit the GCR energy spectral shape for every month,
\begin{center}
\begin{equation}
g(E,t)=\frac{E^{2}+2E_{0}E}{E_{m}^{2}}(\frac{E+E_{0}}{E_{m}})^{\eta (t)}\label{eq12},
 \end{equation}
\end{center}
where $E$ is the kinetic energy in Table \ref{table1}, $E_{0}$ is the rest energy of proton (938.3 MeV), $E_{m}$ is constant (1 GeV), and $\eta (t)$ is the spectral shape parameter for the $t{th}$ month.

\subsection{Energy Spectra Model}
The final GCR energy spectra model can be obtained by combining the integral intensity model $I^{m}(z,t)$ and the spectral shape function $g(E,t)$ according to
\begin{center}
\begin{equation}
f(z,t,E)=I^{m}(z,t)N(z,t)g(E,t)=Z^{m}(z,t)10^{\alpha (t)}N(z,t)g(E,t)\label{eq13},
 \end{equation}
\end{center}
where $f(z,t,E)$ is the modeled differential flux for element $z$ with energy $E$ at time $t$, and $N(z,t)$ is a normalising factor function calculated using Equation \eqref{eq14},
\begin{center}
\begin{equation}
N(z,t)=(\sum_{i=1}^{7} g(E_i,t) \Delta E_{i})^{-1}\label{eq14}.
 \end{equation}
\end{center}

\subsection{ The relation between model parameters and sunspot number}
A practical GCR model should be able to reconstruct GCR spectra for earlier periods and also be capable of predicting the future space radiation environment for manned missions. Using a similar method as in the CR$\grave{E}$ME/Nymmik model, the ZQ13 model connects their model parameters, including the integral intensity modulation parameter $\alpha (t)$, the averaged integral intensity ratio percentage $\bar{p}(t)$, and the spectral shape function parameter $\eta (t)$ with yearly averaged sunspot numbers. Thus the model parameters can be estimated through the SSN record during earlier periods. Considering the propagation time of the dynamic solar wind plasma and the embedded interplanetary magnetic field to the boundary of the heliosphere, the ZQ13 model takes the time lag as 1 year for simplicity. Comparing the reconstructed GCR fluxes from the ZQ13 model with other models (CR$\grave{E}$ME and BON model), and with historical measurements (IMP-8), the ZQ13 model offers obvious advantages in describing GCR spectra.

\begin{figure}[H]
\centering
\includegraphics[width=9cm]{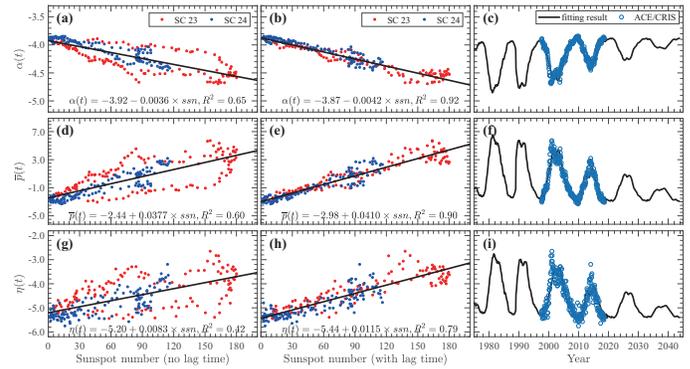}

\caption{Intensity modulation parameter $\alpha (t)$, averaged IRP $\bar{p}(t)$, and spectral shape function parameter $\eta(t)$ obtained by fitting monthly averaged ACE/CRIS observations as a function of sunspot number (red dots for solar cycle 23, and blue dots for solar cycle 24). The solid lines in panels (c), (f), and (i) show the monthly parameters obtained from the fitting functions in panels (b), (e), and (h), and the blue circles are the model parameters calculated from ACE/CRIS measurements. SSNs after 2018 are digitized from Figure 11 of reference \cite{ref46}.}
\label{Fig:Fig.1}
\end{figure}

\begin{figure}[H]
\centering
\includegraphics[width=9cm]{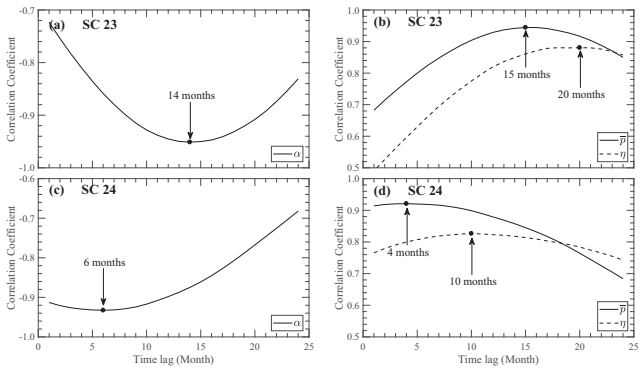}
\caption{The cross-correlation coefficients between model parameters ($\alpha (t)$, $\bar{p}(t)$, and $\eta(t)$) and lagged sunspot number for odd (panels (a) and (b)) and even solar cycles (panels (c) and (d)), respectively. The arrows mark the optimum time lag used in our model.}
\label{Fig:Fig.2}
\end{figure}

Consider the solar activity (represented by SSN) dependence of the model parameters calculated from ACE/CRIS observations. As shown in the first two columns of \cref{Fig:Fig.1}, all parameters display certain correlations with SSN. A negative correlation between $\alpha (t)$ and SSN exists together with positive correlations between $\bar{p}(t)$, $\eta (t)$ and SSNs. Note that SSNs and parameters in the first column ( \cref{Fig:Fig.1} (a), (d) and (g)) assume the same time (i.e., without time lag), while time lags, as determined by \cref{Fig:Fig.2}, are considered in the second column ( \cref{Fig:Fig.1} (b), (e) and (h)). \cite{ref47} confirms that the lag effect must be reflected in the features of GCR spectra, and they \textcolor{black}{also} point out that this effect is different during odd and even solar cycles.

To determine the appropriate time lag of each parameter, we calculate the cross-correlation coefficients between the model parameter and sunspot number with time lags from 1 to 24 months, with odd (SC 23) and even (SC 24) cycles considered separately. The maximum absolute cross-correlation coefficient is determined and the corresponding time lag is regarded as the optimal time lag as marked by the arrows in Figure
\ref{Fig:Fig.2}. For odd cycle, the lagged times for $\alpha (t)$, $\bar{p}(t)$, $\eta (t)$ are 14, 15 and 20 months, respectively, and a corresponding 6, 4 and 10 months are suggested for the even cycle. These lagged months are used in \cref{Fig:Fig.1} (b), (e), and (h). We can see that without a delay (\cref{Fig:Fig.1} (a), (d), and (g)) the linear correlation coefficients are relatively low but are significantly improved when a lagged time is included (\cref{Fig:Fig.1} (b), (e), and (h)). \textcolor{black}{Therefore,} a reasonable delay is required to improve GCR model accuracy. With the lagged time and the linear fitting functions (noted in \cref{Fig:Fig.1} (b), (e), and (h)), model parameters are built using historical or future predictive SSNs (\cref{Fig:Fig.1} (c), (f) and (i)). The blue circles in (c), (f) and (i) are the model parameters calculated from ACE/CRIS. \textcolor{black}{The modeled results and ACE/CRIS measurements show a good agreement}.

\section{RESULTS}\label{sec:4}
Theoretically, with the model described in Section 3, we can obtain the GCR flux for any element with a certain energy at any time of interest, and the lagged sunspot number is the unique input parameter. In this section, we compare GCR fluxes calculated from our model with those from measurements (ACE/CRIS and SIS, IMP-8), and also with those from other models (CR$\grave{E}$ME2009, BON2010, 2011 and 2014). In order to evaluate the accuracy of GCR model, the relative difference $Rd$ between the modeled spectra at 1 AU and the corresponding measurements is
\begin{center}
\begin{equation}
Rd=\frac{1}{M}\sum_{k=1}^{M}\frac{Model_{k}-Obs_{k}}{Obs_{k}}\times 100 \%\label{eq15},
 \end{equation}
\end{center}
\textcolor{black}{where} $M$ is the number of measurements. Equation \eqref{eq15} has been applied in part to previous validation studies, such as references \cite{ref39}, \cite{ref44}, \cite{ref18} and \cite{ref41}. \textcolor{black}{We note} that the total mean of $Rd$ cannot provide a real description of model discrepancy because of the positive-negative cancellation effect. The absolute relative difference $|Rd|$ is applied \textcolor{black}{to decide} the overall model accuracy, which is
\begin{center}
\begin{equation}
|Rd|=\frac{1}{M}\sum_{k=1}^{M}\frac{|Model_{k}-Obs_{k}|}{Obs_{k}}\times 100 \%\label{eq16}.
 \end{equation}
\end{center}

Additionally, it should be mentioned that GCR measurements are usually discontinuous over the energy range, while modeled results are continuous in energy. In order to compare model results with measurements, we integrate the model output (our model and CR$\grave{E}$ME model) to match the same measured energy bin. There is no obvious difference if the comparison is done by directly using the recommended midpoint of each energy bin instead (as listed in Table \ref{table1}).

\subsection{Integral Intensity of GCR Particles}
\begin{figure}[H]
\centering
\includegraphics[width=8cm]{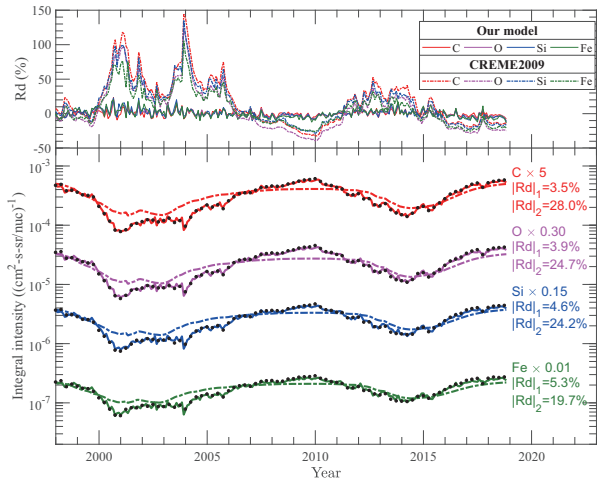}
\caption{Measured and modeled integral intensities of GCR elements: carbon (red), oxygen (magenta), silicon (blue) and iron (green). The solid lines represent our model results and the dashed lines are the CR$\grave{E}$ME2009 model \textcolor{black}{results}. The corresponding relative differences are shown in the top panel. The black solid cycles are measurements from ACE/CRIS. $|Rd|_1$ \textcolor{black}{denotes} our model, and $|Rd|_2$ the CR$\grave{E}$ME2009 model.}
\label{Fig:Fig.3}
\end{figure}

GCR energy spectra for carbon, oxygen, silicon, and iron nuclei measurements obtained from ACE/CRIS together with the corresponding spectra provided by our model and the CR$\grave{E}$ME2009 model are investigated. The monthly integral intensities from January 1998 \textcolor{black}{to} October 2018 are presented in \cref{Fig:Fig.3}.

It is evident that the intensity derived from our model is significantly better \textcolor{black}{than} that described by the CR$\grave{E}$ME2009 for these four nuclei. Our model fits the measurements very well, while the CR$\grave{E}$ME2009 model exhibits large discrepancies. For our model, the absolute differences ($|Rd|$) are 3.5, 3.9, 4.6, and 5.3\% for carbon, oxygen, silicon, and iron nuclei, respectively, and the corresponding values are 28.0, 24.7, 24.2, and 19.7\% for the CR$\grave{E}$ME2009 model.

\subsection{Differential Flux of GCR Particles}
\begin{figure}[H]
\centering
\includegraphics[width=8cm]{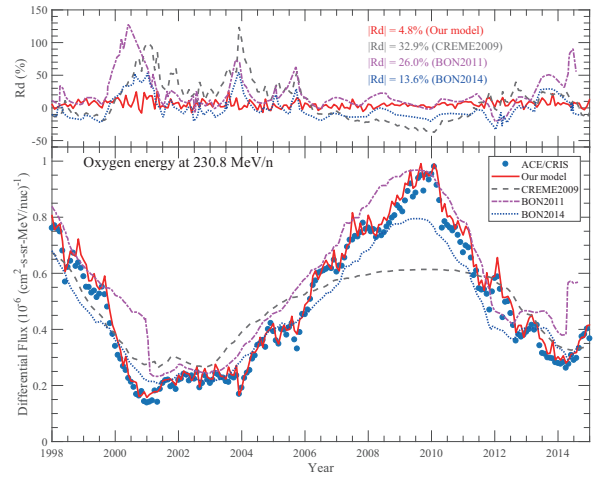}
\caption{Measured and modeled differential flux for oxygen with energy 230.8 MeV/nuc. The model relative differences are plotted in the top panel. BON model results are digitized from Figure 8 of reference \cite{ref41}.}
\label{Fig:Fig.4}
\end{figure}

\cref{Fig:Fig.4} shows the monthly differential flux for oxygen with energy 230.8 MeV/nuc. The fluxes for the BON2011 and BON2014 model were digitized from reference \cite{ref41}. Measurements from ACE/CRIS are also plotted for the purpose of evaluating the model accuracy. The CR$\grave{E}$ME2009 model significantly underestimates the measured fluxes between 2006 and 2012, and the BON2011 model overestimates the measurements for most of the time of interest. BON2014 and our model, by contrast, is in good accord with the measurements. The averaged absolute differences \textcolor{black}{in} our model, CR$\grave{E}$ME2009, BON2011 and BON2014 are 4.8, 32.9, 26.0, and 13.6\%, respectively, indicating that our model provides the best description for the differential flux of oxygen.

\begin{figure}[H]
\centering
\includegraphics[width=8.2cm]{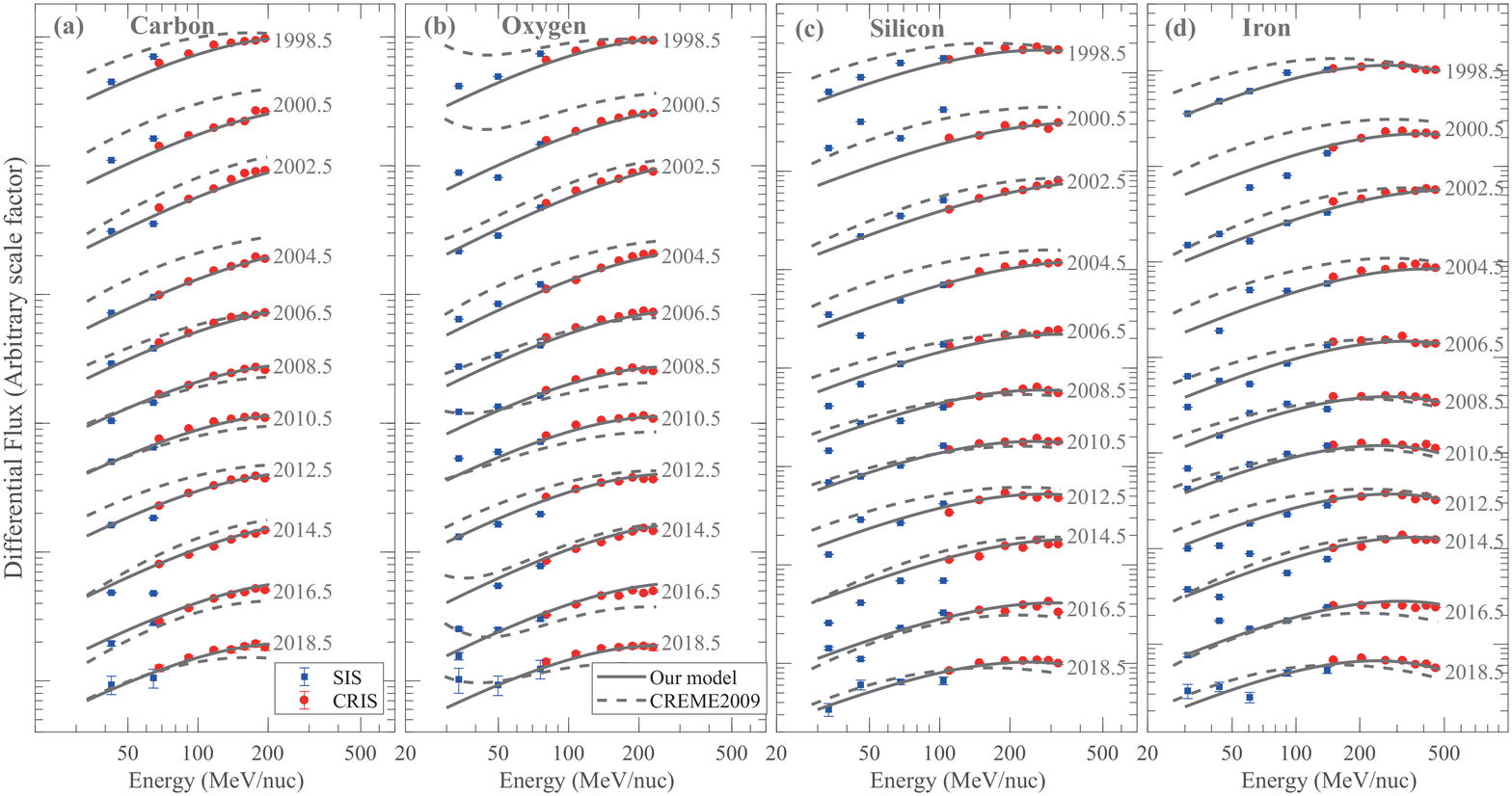}
\caption{Measured and modeled annual GCR energy spectra for carbon (a), oxygen (b), silicon (c) and iron (d) from June 1998 to June 2018. The solid lines are our model results, and the dashed lines for CR$\grave{E}$ME2009. Dots with error bars are the measurements from ACE/CRIS and SIS.}
\label{Fig:Fig.5}
\end{figure}

\cref{Fig:Fig.5} shows the GCR energy spectra during the period of 1998-2018 (even-numbered years only) described by our model, CR$\grave{E}$ME2009 model, and ACE/CRIS measurements. Elements carbon (a), oxygen (b), silicon (c) and iron (d) are studied. \cref{Fig:Fig.5} shows that our modeled results agree with the measurements very well for most of the time and almost \textcolor{black}{every} energies. The CR$\grave{E}$ME2009 model, however, does not yield such a good fit. The annual relative differences for the two models are listed in Table \ref{table3}. It can be seen that more that 75\% of $Rd$ for CR$\grave{E}$ME2009 model are positive, which means that the model overestimates the GCR flux in the ACE measured energy range, and its overall difference and absolute difference is 19.2 and 29.6\%. The $Rd$ of our model varies from $-7.2 \sim 17.0\%$ (the overall is 2.8\%), and the absolute relative difference ($|Rd|$) from $9.3\%$ to $21.6\%$ (the overall is 14.0\%). In summary, the CR$\grave{E}$ME2009 model systematically overestimates the GCR measurement, whereas our model presents a more balanced prediction.

\begin{table}[H]
\centering
   \tabcolsep 5pt 
   \footnotesize
  \caption{Annual relative difference for all elements}
  \label{tab:performance_comparison}
    \begin{tabular}{ccccc}
    \toprule
    \multirow{2}{*}{Year}&
    \multicolumn{2}{c}{$Rd$ (\%)}&\multicolumn{2}{c}{$|Rd|$ (\%)}\cr
    \cmidrule(lr){2-3} \cmidrule(lr){4-5}
    &Our model&CR$\grave{E}$ME2009&Our model&CR$\grave{E}$ME2009\cr
    \midrule
    1998.5 & -1.3  & 24.2  & 10.1  & 27.1 \\
    1999.5 & -0.1  & 1.2   & 13.3  & 15.7 \\
    2000.5 & -0.3  & 56.8  & 16.5  & 59.4 \\
    2001.5 & 7.2   & 39.8  & 18.3  & 42.8 \\
    2002.5 & 0.2   & 24.3  & 13.6  & 29.9 \\
    2003.5 & 12.0    & 68.5  & 21.6  & 69.4 \\
    2004.5 & 0.5   & 45.9  & 13.0    & 47.6 \\
    2005.5 & -1.0    & 44.5  & 14.3  & 46.1 \\
    2006.5 & -2.7  & 10.0    & 11.2  & 16.3 \\
    2007.5 & 11.5  & 1.6   & 14.7  & 13.6 \\
    2008.5 & 3.1   & 4.0     & 11.2  & 16.4 \\
    2009.5 & -7.2  & -16.2 & 10.8  & 20.4 \\
	2010.5 & -1.1  & -2.2  & 11.3  & 16.2 \\
	2011.5 & 3.3   & 26.5  & 13.7  & 29.4 \\
	2012.5 & 6.1   & 30.8  & 12.1  & 31.8 \\
    2013.5 & 17.0    & 37.4  & 20.7  & 38.5 \\
	2014.5 & 9.2   & 21.4  & 14.8  & 25.7 \\
	2015.5 & 1.9   & 9.9   & 15.8  & 19.7 \\
	2016.5 & 13.5  & -7.7  & 15.9  & 18.3 \\
	2017.5 & -8.6  & -10.7 & 13.1  & 20.2 \\
	2018.5 & -2.8  & -6.5  & 9.3   & 16.7 \\
	\bfseries Mean  & \bfseries 2.8   & \bfseries 19.2  & \bfseries 14.0    & \bfseries 29.6 \\
	\bottomrule
    \end{tabular}
   \label{table3}%
\end{table}

\begin{figure}[H]
\centering
\includegraphics[width=8cm]{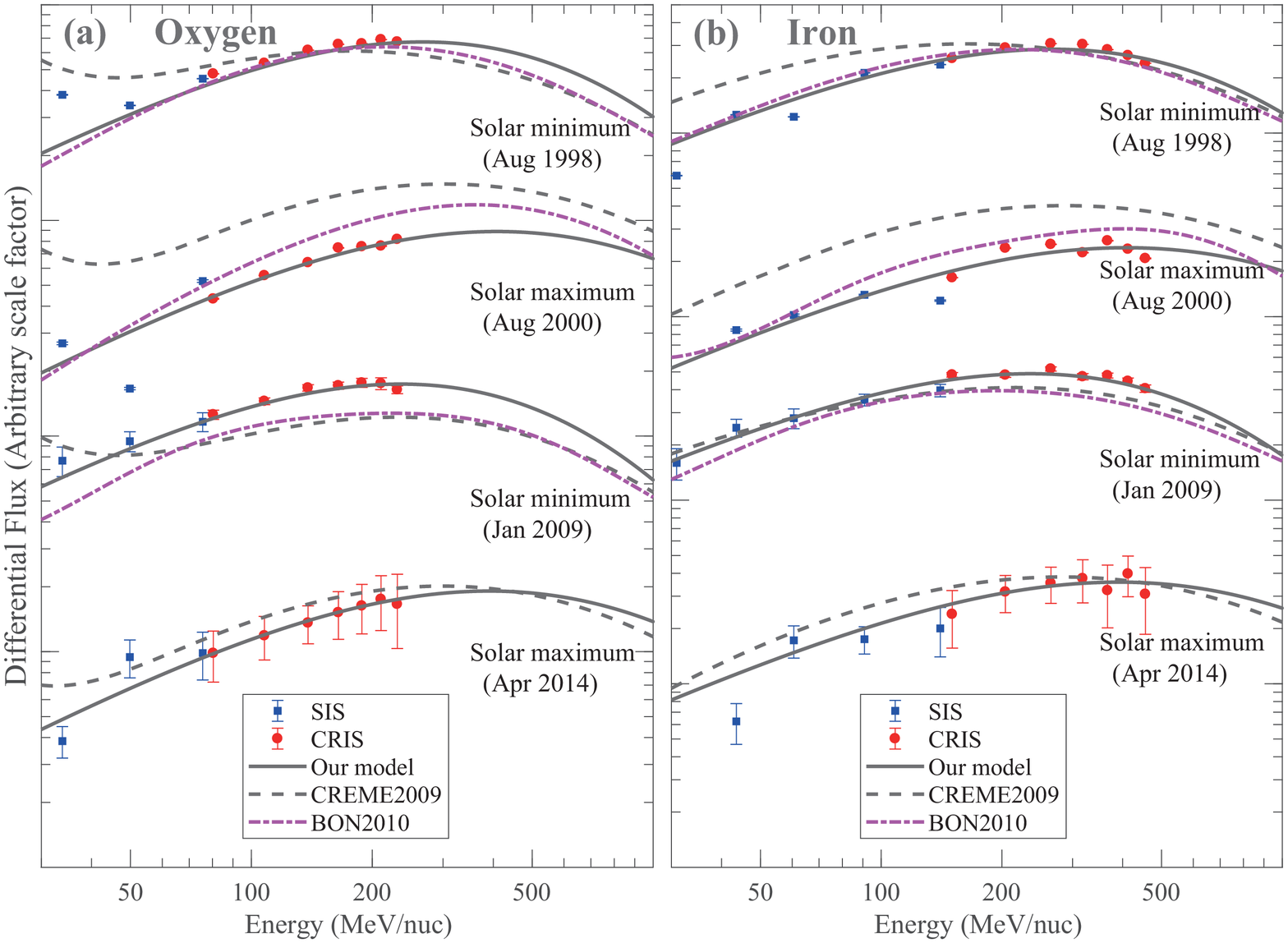}
\caption{Measured and modeled GCR energy spectra under different solar modulation conditions, (a) Oxygen, (b) Iron. The BON2010 model is digitized from Figure 2 of reference \cite{ref44}.}
\label{Fig:Fig.6}
\end{figure}

\cref{Fig:Fig.6} shows modeled GCR energy spectra during the recent two solar minima (August 1998 and January 2009) and two maxima (August 2000 and April 2014). Representative elements are oxygen (a) and iron (b). The corresponding BON2010-derived fluxes were digitized from reference \cite{ref44}, and they select the date on the basis of Bartels Rotations (BR). For example, they show spectra over the period of 31 July to 26 August (BR\# 2253) in 1998 to represent the first solar minimum condition. To compare with their work, we approximately use August 1998 as the corresponding time and show the experimental and \textcolor{black}{model} results. Such a small time interval does not bring large discrepancies. BON modeled results are not available for April 2014. From Figure \ref{Fig:Fig.6}, we can see that, in 1998, our model and the CR$\grave{E}$ME2009 model show good fitting results for these two elements, but for other time periods, such as, August 2000, January 2009, April 2014, only our model always provides relatively good agreement with the ACE/CRIS and ACE/SIS observations.

\subsection{Statistics of the Model Accuracy}

\begin{figure}[H]
\centering
\includegraphics[width=8cm]{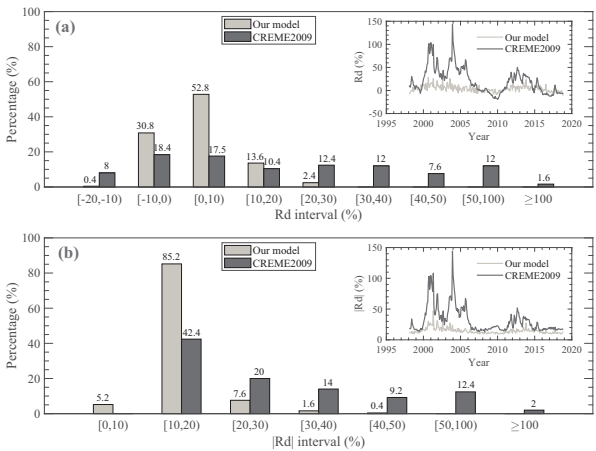}
\caption{The statistical percentage distribution of the binned relative difference $R_d$ (a) and its absolute values $|R_d|$ (b) for our model (grey) and CR$\grave{E}$ME2009 (black), for modeling monthly GCR energy spectra between January 1998 and October 2018.}
\label{Fig:Fig.7}
\end{figure}

We predict monthly GCR energy spectra for all months (January 1998-October 2018) and all nuclei (z=5 to 28), and calculate the relative differences between model outputs and measurements. The relative differences ($Rd$) are binned into 9 intervals, i.e., [-20, -10\%), [-10, 0\%), [0, 10\%), [10, 20\%), [20, 30\%), [30, 40\%), [40, 50\%), [50, 100\%), [100, 150\%), and the absolute relative differences ($|Rd|$) are grouped into 7 intervals [0, 10\%), [10, 20\%), [20, 30\%), [30, 40\%), [40, 50\%),[50, 100\%), [100, 150\%). We count the percentage in each interval and show the results in Figure \ref{Fig:Fig.7}. The panels in the upper right corner exhibit the variations of the monthly differences over time.

In \cref{Fig:Fig.7}(a), more than 80\% of the $Rd$ values \textcolor{black}{of} our model \textcolor{black}{lies} in the range of -10$\sim$10\%, and there is no $Rd$ greater than 30\%. For CR$\grave{E}$ME2009, the $Rd$ values are also mainly around 10\% but they are more scattered, with 33.2\% of $Rd$ values greater than 30\% and 13.6\% greater than 50\%. In \cref{Fig:Fig.7}(b), most $|Rd|$ values concentrate in the interval 10$\sim$20\% for both our model and CR$\grave{E}$ME2009. Some, 5.2\% of $|Rd|$ values are less than 10\% \textcolor{black}{in} our model (there is no corresponding data for CR$\grave{E}$ME2009), while 14.4\% of $|Rd|$ values are greater than 50\% for CR$\grave{E}$ME2009 model (no corresponding data for our model). Furthermore, $Rd$ and $|Rd|$ of CR$\grave{E}$ME2009 exhibit a positive correlation with solar activity, meaning that a higher discrepancy occurs near solar maximum and a smaller discrepancy near solar minimum.

\begin{figure}[H]
\centering
\includegraphics[width=8cm]{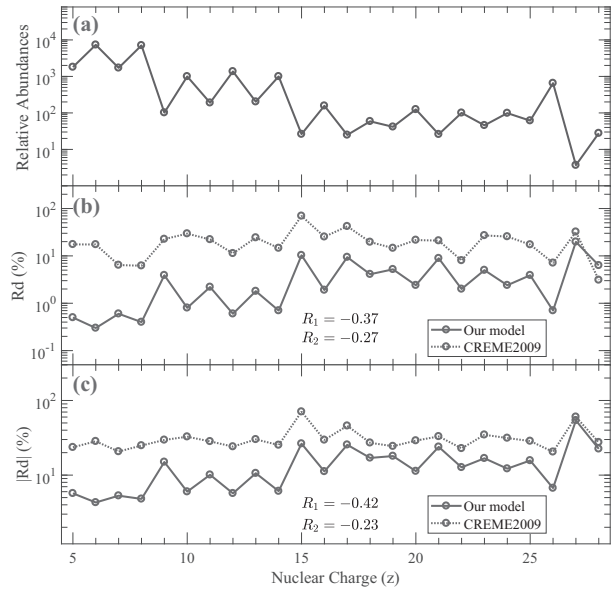}
\caption{(a) The GCR solar minimum elemental relative abundances from \cite{ref48}. (b) Relative difference varies with nuclear charge $z$. (c) Absolute relative difference varies with nuclear charge $z$. In panels (b) and (c), $R_1$ is the correlation coefficient between our model $Rd$ (or $|Rd|$) and elemental relative abundances, and $R_2$ between the CR$\grave{E}$ME2009 model and elemental relative abundance.}
\label{Fig:Fig.8}
\end{figure}

\cref{Fig:Fig.8} shows the elemental dependence of the relative difference. The ACE/CRIS solar minimum abundances are reported in reference \cite{ref48}. By calculating the linear correlation between the modeling relative difference ($Rd$ and $|Rd|$) and the relative elemental abundances, a negative correlation is found, which means that the larger the elemental abundance is, the better the modeling performance is, and vice versa. Additionally, our model exhibits a higher correlation with elemental abundance than CR$\grave{E}$ME2009.

Table \ref{table4} lists the overall averaged relative difference, including $Rd$ and $|Rd|$, \textcolor{black}{of} our model, CR$\grave{E}$ME2009, BON2011 and BON2014. For our model and the CR$\grave{E}$ME2009 model, the values are determined by averaging the relative difference over all elements and all months from January 1998 to October 2018. For the BON model, values are directly taken from reference \cite{ref41}. As shown in Table \ref{table4}, the $Rd$ values \textcolor{black}{of} our model, CR$\grave{E}$ME2009, BON2011 and BON2014 are 3.9, 20.8, 17.9, and -0.4, respectively, and the corresponding $|Rd|$ values are 14.5, 31.4, 23.7 and 13.0. Considering the positive-negative cancellation effect in $Rd$, the BON2014 model provides a more balanced GCR prediction than the other three models which uniformly overestimate GCRs. The $|Rd|$ value of our model is 14.5\%, which is slightly higher than the BON2014 model. From the above analysis, we are confident \textcolor{black}{that our model predicts GCR spectra well}.

\begin{table}[H]
  \caption{The averaged relative difference between modeled results and measurements}
  \centering
   \tabcolsep 18pt 
   \footnotesize
    \begin{tabular}{rrr}
    \toprule
    Model & $Rd$ (\%) & $|Rd|$ (\%) \\
    \midrule
    Our model & 3.9   & 14.5 \\
    CR$\grave{E}$ME2009 & 20.8  & 31.4 \\
    BON2011 & 17.9  & 23.7 \\
    BON2014 & -0.4  & 13.0 \\
    \bottomrule
    \end{tabular}%
  \label{table4}%
\end{table}%

\subsection{Predicted GCR Flux}
The empirical formulas \textcolor{black}{that describe} the relationship between the model parameters and the monthly mean sunspot number are well established, and the accuracy of our model in predicting \textcolor{black}{the} GCR spectra over a long-term scale has been carried out \cite{ref1}. Consequently, we now have the capability to present GCR spectra for any time of interest (the temporal resolution is one month) and element ($z$ = 5 to 28) as long as the SSN is given. The accuracy of predicting the upcoming GCR environment depends in part then on the forecast of the SSN, but that is not the core issue of this paper. Many techniques and methods have been proposed to forecast the SSN over the years \cite{ref49, ref50, ref51}. Here, we illustrate the applicability of our GCR model and use it to derive a predictable GCR environment for the readers. In this paper, the prediction of SSNs for the following two solar cycles is extracted from reference \cite{ref46}).

\begin{figure}[H]
\centering
\includegraphics[width=8.5cm]{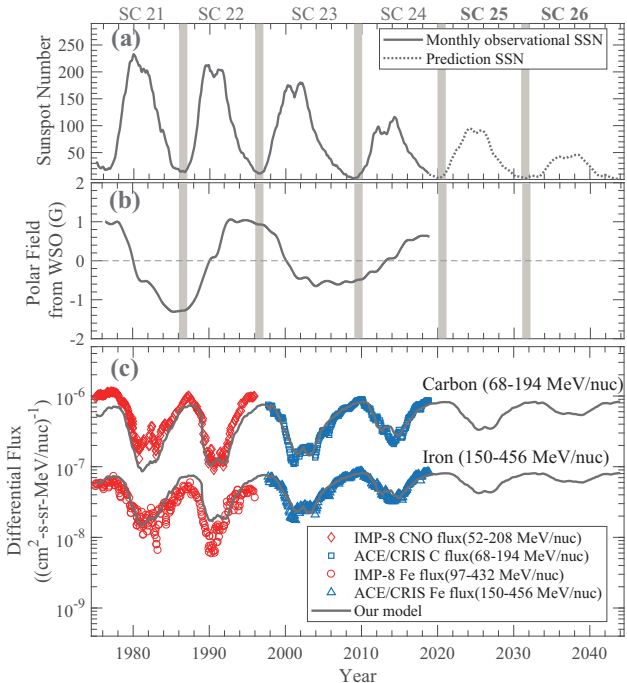}
\caption{(a) Sunspot number during 1975-2044. The solid line shows the historical observations from SILSO (\url{http://www.sidc.be/silso/datafiles}), and the dashed line is the predicted sunspot number in the following two solar cycles digitized from reference \cite{ref46}. (b) 10-day average of solar polar field strength (North and South) from the WSO site (\url{http://wso.stanford.edu/Polar.html}) from January 1975-November 2018. Use of a 20 nHz low pass filtered values eliminates yearly geometric projection effects. (c) The modeled carbon and iron differential fluxes (solid line), and the observations (scattered points) from IMP-8 digitized from reference \cite{ref35} and from ACE/CRIS.}
\label{Fig:Fig.9}
\end{figure}

\cref{Fig:Fig.9}(a) shows SSNs during 1975-2044. \cref{Fig:Fig.9}(b) is the 10-day averaged solar polar field strength. The polarity of the solar polar field is supposed to influence the time lag between solar activity and GCR intensity at 1 AU \cite{ref47}. \cref{Fig:Fig.9}(c) displays the predictions of the GCR differential fluxes from 1975 to 2044 for carbon and iron nuclei. The corresponding measurements from IMP-8 and ACE/CRIS are displayed as a scatter plot. Before 1996, the IMP-8 data is digitized from reference \cite{ref35}. It should be noted that the energy interval of modeled iron is not perfectly consistent with that of IMP-8, and we present the carbon flux instead of CNO. \textcolor{black}{O}ur modeled results agree well with the early IMP-8 measurements in trend. After 1998, the predictions show good consistency with ACE/CRIS observations in terms of variation and magnitude. For the following two solar cycles, the GCR flux does not show too much difference between \textcolor{black}{the} solar maximum and minimum which is thought to be due to the weaker solar cycles. Note that our GCR prediction here can only serve as a reference. To a certain extent, more accurate GCR forecasts rely on the accurate prediction of SSNs.

\section{Discussion and Conclusion}\label{sec:5}
Based on yearly averaged GCR heavy nuclei measurements from the ACE/CRIS instrument, Zhao and Qin proposed an empirical and phenomenological model to describe GCR energy spectra over the energy range 30-500 MeV/nuc \cite{ref1}. The model predictions are consistent with measurements obtained from either ACE or IMP-8 spacecraft, showing the validity of the model. However, the ZQ13 model can only provide annual mean GCR fluxes which cannot assess the short-term GCR intensity. The aim of this paper is to build an empirical GCR model with higher time precision and more accurate prediction on the basis of the ZQ13 model.

(1) The latest (until October, 2018) highly statistically precise GCR flux measurements from ACE/CRIS, covering nearly two solar cycles, are collected and used in developing our model.

(2) Our model is characterized by three key parameters: the intensity modulation parameter $\alpha(t)$, the averaged intensity ratio percentage (IRP) parameter $\bar{p}(t)$, and the spectral shape function parameter $\eta(t)$. By fitting the latest monthly ACE/CRIS measurements, our model can now provide monthly averaged GCR spectra for heavy nuclei at any time of interest, which is a notable improvement over the ZQ13 model.

(3) The ZQ13 model is built on two assumptions, the first being that all elements share the same spectral shape at the same time, and the second that the intensity ratio percentage (IRP) for all elements is generally the same (equals to 1), excluding C(6), O(8) and Fe(26). The former is still applied in the current work, while the latter is no longer adopted. The refined parameter $\lambda^*(z)$, which is used to calculate the IRP of each element, now has 24 choices corresponding to 24 elements ($5 \leq z \leq 28$).

(4) Relatively reasonable and accurate time lags for odd and even solar cycles are included in our model. To include the hysteresis of solar modulation, an approximate one-year time lag between sunspot number record and model parameters is taken into account in the ZQ13 model. However, as documented in previous literature, the effects of solar activity on GCR modulation should be different for odd and even solar cycles \cite{ref4, ref13}. \cite{ref46}, for example, use a 15.5-months lag for odd cycles and a 5.5-months lag for even cycles. By calculating the cross-correlation coefficients between  our model parameters and the lagged sunspot numbers (lag time ranges from 1 to 24 months), the time lag corresponding to the maximum absolute coefficient is chosen for our model. For odd solar cycles, the time lag used in our model is 14, 15, and 20 months for parameters $\alpha(t)$, $\bar{p}(t)$, and $\eta(t)$ respectively, while for even cycles they are 6, 4, and 10 months accordingly. We show that our model is able to reproduce elemental GCR spectra during 1975 to 2018, showing good agreement with the observations from IMP-8 and ACE/CRIS.

(5) The model accuracy was performed by statistically analysing the relative differences between the modeled results and measurements. Also, we compared our modeled results with other well-known GCR models, such as the CR$\grave{E}$ME2009, BON2010, 2011 and 2014 models. Our model shows much better consistency with the observations than the CR$\grave{E}$ME2009 model for the energy range 30-500 MeV/nuc and has a comparable prediction accuracy with the BON2014 model.

(6) Future near-Earth GCR carbon and iron nuclei fluxes are predicted for the following two solar cycles. The peak intensities for both nuclei decrease gradually with time, while the valley intensities increase gradually. The amplitude of flux variation decreases with time.

\Acknowledgements{This research work is partly supported by the National Science Foundation of China (41174165, 41504133), L.L.Z and G.P.Z
acknowledge the partial support of the NSF EPSCoR RII-Track-1 Cooperative Agreement OIA-1655280,
NASA grants NNX08AJ33G, Subaward 37102-2, NNX14AC08G, NNX14AJ53G, A99132BT, RR185-447/4944336 and NNX12AB30G. S.F. and Y.J. acknowledge partial support of National Key R\&D Program of China (2018YFC1407304, 2018YFF01013706),  Open Fund of Key Laboratory (201801003), and other Foundation (315030409). The data sets of SIS and CRIS are download from the ACE Science Center archives at \url{http://www.srl.caltech.edu/ACE/ASC/level2/}. The monthly averaged sunspot numbers are download from \url{http://www.sidc.be/silso/datafiles}. The solar polar field strength (North and South) is obtained from the Wilcox Solar Observatory (WSO) site at \url{http://wso.stanford.edu/Polar.html}. The CR$\grave{E}$ME2009 model predictions are produced by CR$\grave{E}$ME site (\url{https://creme.isde.vanderbilt.edu/}).}

\InterestConflict{The authors declare that they have no conflict of interest.}



\end{multicols}

\end{document}